\documentclass{article}
\usepackage{graphicx}
\begin{document}

\begin{center}
{\bf \large
Long-tail Feature of 
DNA Words Over- and Under-representation in Coding Sequences}
\end{center}

\vspace*{1cm}
\begin{flushleft}
{\large (runtitle: Long-tail Feature of DNA Words)}
\end{flushleft}

\vspace*{2cm}
\begin{center}
{A. Nowicka$^1$, M.R. Dudek$^2$, S. Cebrat$^1$, M. Kowalczuk$^1$, 
 P. Mackiewicz$^1$, M. Dudkiewicz$^1$, D. Szczepanik$^1$}\\
~\\
{$^1$ \it Institute of Microbiology, University of Wroc{\l}aw,
ul. Przybyszewskiego 63/77\\ 
54-148 Wroc{\l}aw, Poland}
~\\
~\\
{$^2$ \it Institute of Theoretical Physics, University of Wroc{\l}aw,
        pl. Maxa Borna 9\\
 50-204 Wroc{\l}aw, Poland
}
\end{center}

\begin{abstract}
We have analyzed DNA sequences of known genes from 16 yeast chromosomes 
({\it Saccharomyces cerevisiae}) 
in terms of oligonucleotides.   We have noticed that the relative abundances 
of oligonucleotide usage in the genome follow a long-tail L\'{e}vy-like 
distribution. 
 We have observed that long genes  often 
use strongly over-represented and under-represented nucleotides,
whereas it was not the case for the short genes (shorter than 300 nucleotides) 
under consideration. 
 If selection on the extremely over-represented/under-represented
oligonucleotides  
was strong, long genes would be more affected by 
spontaneous mutations 
than short ones. 
\end{abstract}


\pagebreak
\section{Introduction}
Since 1995 more than  thirty full genomes have been sequenced and
information on the sequences of hundreds of millions of
nucleotides has been available for the scientific community. This
has opened a new field of research, DNA statistical analysis,  where genomic
sequences are analyzed both in terms of single nucleotides,  
oligonucleotides and thousands of
nucleotides. In the last case there are studies on the 
power spectral density and correlation function, especially the question
of the existence of statistical long-range
base-base correlation.
Long-range correlation in DNA 
was first observed in 1992 by three groups,
Li et al. \cite{l_Li0,l_Li1}, Peng et al. \cite{l_Stan0} and
Voss \cite{l_Voss1}. This has been a very active topic until now. We
will not put a long list of references here but we cite 
only some of the 
the recent papers by
H.E. Stanley et al. \cite{l_stan1}, Arn\'{e}odo et al.
\cite{l_Arneodo2} and  Vieira \cite{l_vieira} addressing the
problem directly. One can also visit a WWW home page by Li
({\it http://linkage.rockefeller.edu/wli/dna\_corr}) 
for the references to this particular topic.
We have also contributed to the topic, e.g., in the papers 
\cite{l_cebrat}-\cite{l_Mackiewicz2} we show that
replication, which is an asymmetric process, is responsible for
introducing strong trends
in the third bases of codons and in consequence it causes the long-range
base-base correlations.

The examinination of the long-range correlation in DNA is strongly
connected with the statistical 
methods applied to texts in natural languages
\cite{l_Ebeling}-\cite{l_ausloos2}, 
where usually one calculates the frequency $f(k)$ of each word in a
text ($k=1,2, \ldots, N$). If the words in the text are arranged in rank order, from
most frequent to least frequent, so that $f(1) \ge f(2) \ge f(3)
\ldots \ge f(N)$ then one observes a power law (Zipf law),   
$f(k) \propto 1/k^{\zeta}$, with an exponent $\zeta$ and typically
$\zeta \sim 1.0$ for natural languages. This analysis also
applies well to studying short-range time correlations in financial
signals \cite{l_ausloos1,l_ausloos2}. In DNA the words are
composed from an "alphabet" of four letters $A$,$T$,$G$,$C$
representing the nucleotides {\it adenine, thymine, guanine} and
{\it cytosine}. The $n$-tuplets are termed "n-words".
The biological meaning of these $n$-words depends on the value of
$n$. Typically in the
case of coding DNA sequences the words are considered to be $3$-tuplets  
because three nucleotides (codons)  code for
one amino acid. This triplet structure of DNA coding sequences
can be easily detected with the help  of the power spectrum because there is
a sharp peak at frequency $f=1/3$ in the spectrum. The connection of the
peak with the codon structure has been reported already by Voss
\cite{l_Voss1} in 1992 during discussion of the long-range
correlations in DNA. This peak reflects the asymmetry of codons.
For example, in the case of the yeast genome ({\it Saccharomyces cerevisiae}) 
more then $75\%$ of
all genes have more $A$ than $T$ in the first and second positions
in codons, more $G$ than $C$ in the first positions of codons, and
less $G$ than $C$ in the second positions \cite{l_cebrat2}. Codons
for hydrophobic amino acids are rich in $T$ in the second
positions whereas codons for hydrophilic amino acids are
rich in $A$ in the second positions. 
In
particular, the genes with lower number of $A$ than $T$ in the
second positions in codons represent genes coding for
transmembrane proteins. Thus, considering  $3$-tuplets in coding
regions to
be the words in DNA texts  is quite natural, 
contrary to the words for
noncoding regions which are not known. On the other hand,  
the observation of a much smaller peak around
$f \sim 1/11$ in DNA power spectrum \cite{l_stan4},\cite{l_herzel} 
makes the understanding the DNA words more complex. Namely, this peak
 might be related to DNA folding structure. The detailed
 discussion of the meaning of the peak can be found in a paper by Trifonov
 \cite{l_trifonov}  as well as a discussion of other
 recognized periodicities in genome sequences, $200-$ and $400-$base
 periodicities.

The results of Zipf analysis
of 40 DNA sequences have been discussed in detail by Mantegna et
al. \cite{l_stan3}. They found that the Zipf exponent $\zeta$ for
a noncoding region is about $50\%$ larger than that for
coding regions and thus noncoding sequences are closer to natural
languages with respect to their information content  
than the coding ones. Note however that also noncoding regions of
DNA can possess a strong signal $f=1/3$ \cite{l_gierlik}.
The reason is that there can be found both
sequences which were coding in past and 
sequences which may be recognized as genes in future. 

\begin{figure}
\begin{picture}(190,130)(5,-30.0)
\put(4,120){G}
\put(105,120){T}
\put(4,-15){A}
\put(105,-15){C}
\put(124,120){G}
\put(226,120){T}
\put(124,-15){A}
\put(226,-15){C}
\put(244,120){G}
\put(356,120){T}
\put(244,-15){A}
\put(356,-15){C}
\includegraphics[height=4cm]{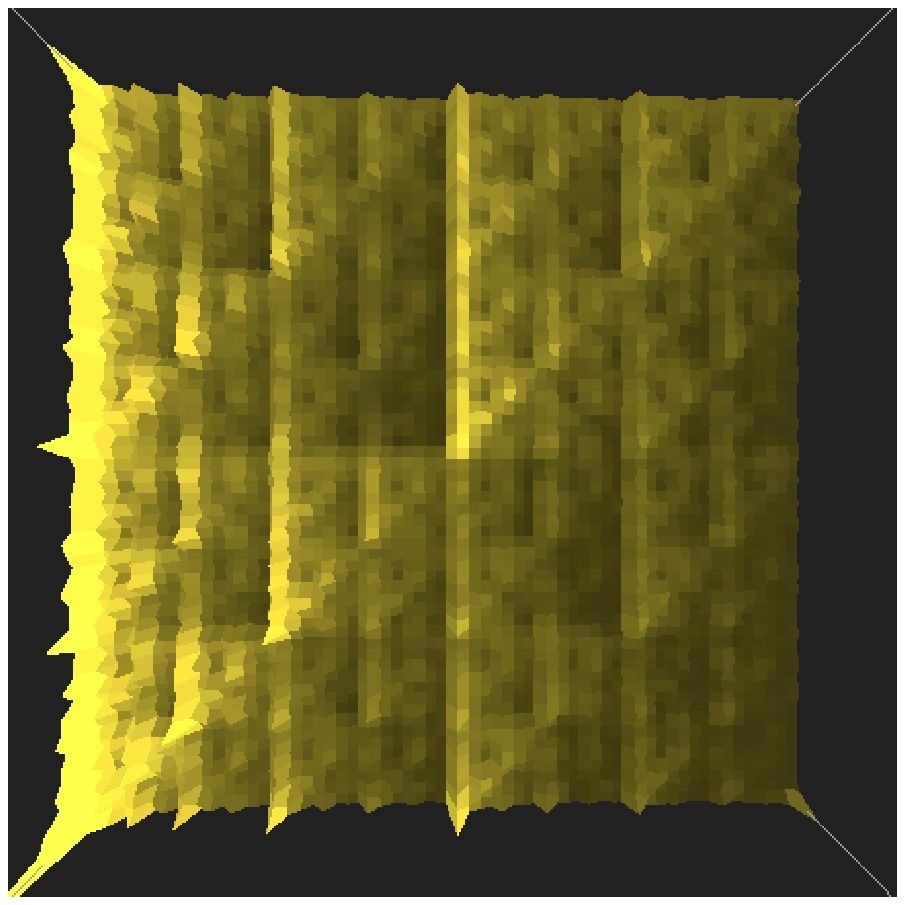}
\hspace{0.1cm}
\includegraphics[height=4cm]{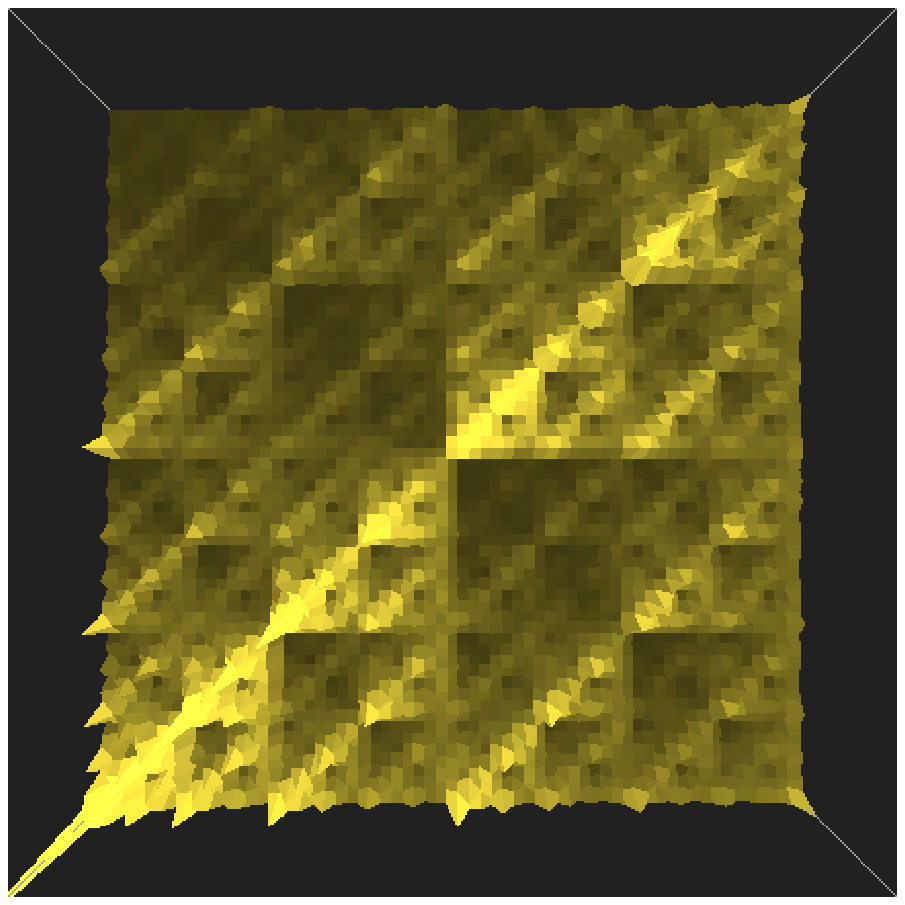}
\hspace{0.1cm}
\includegraphics[height=4cm]{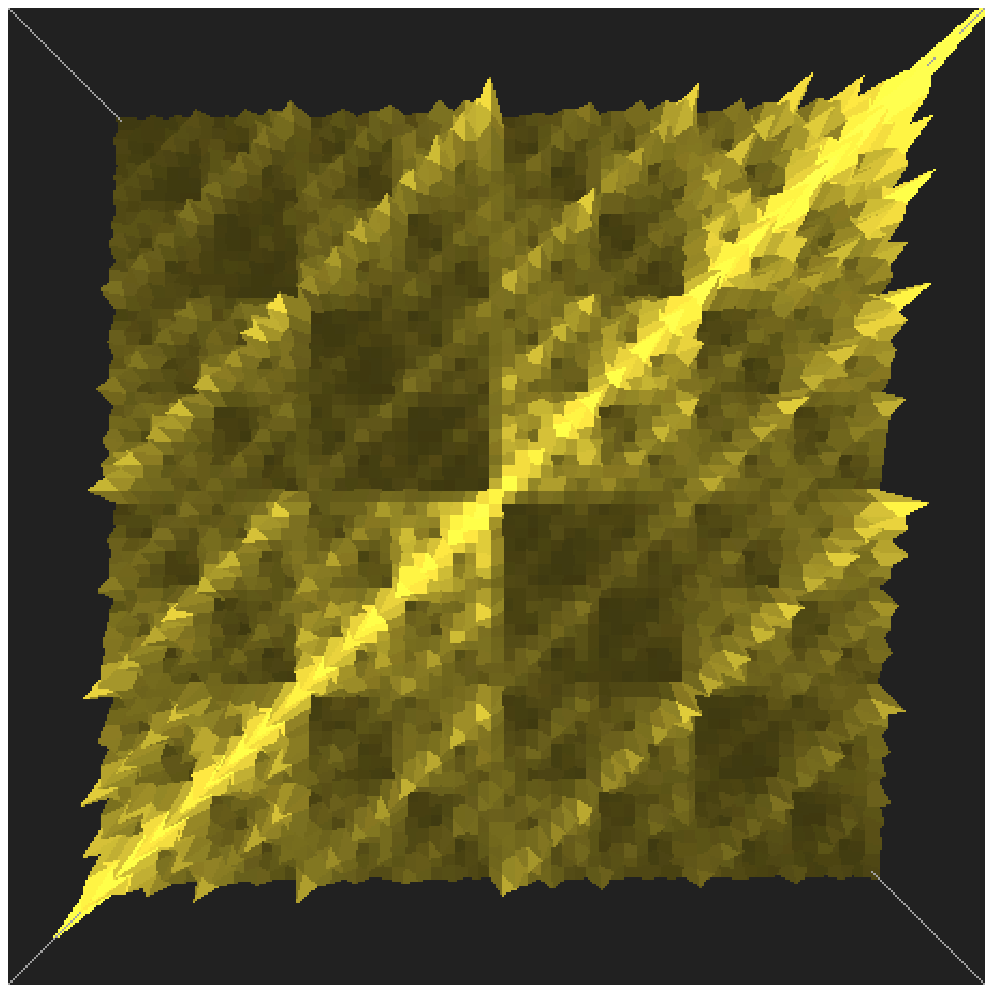}
\end{picture}
\caption{The chaos game representation of the  yeast genes 
 in case of
$6$-words constructed from the first base position in codons (left),
second base position in codons (middle), third base position in codons (right). 
}
\label{fig1}
\end{figure}

The studies of oligonucleotides ($n$-words) in recent years indicate that they can 
play the role of a genomic signature. Karlin and Burge, in their
paper \cite{l_karlin} showed that the relative abundance of
dinucleotides ($2$-words) can discriminate DNA sequences of different
organisms. The abundances, particularly for $CG$ and $TA$
can reflect the species-specific replication and
repair mechanisms (see also Karlin, Mr\`{a}zek and Campbell
\cite{l_karlin2}).  They analyzed different dinucleotides with the help
of effective frequencies:

\begin{equation}
z=\frac{P_{ij}}{P_i P_j}
\label{r_Karlin}
\end{equation}

\noindent
where $P_i$ denoted the frequency of nucleotide $i$ ($i,j=A,T,G,C$)
and $P_{ij}$ denoted the frequency of dinucleotide $ij$ under
consideration. In particular, they suggested that $CG$
under-representation should be advantageous for organisms which
have small genomes and need to replicate rapidly. On the other
hand $TA$ under-representation renders DNA more flexible
for unwinding.    
The concept of genomic signature has been been extended recently
to $n$-words \cite{l_signature} where the chaos game
representation of DNA sequences in the form of fractal images 
has been used following the method developed by Jeffrey
\cite{l_jeffrey}. We address this paper because we have introduced a similar
concept of DNA representation independently in
our paper \cite{l_kowal}. In the method, the frequencies of $n$-words
are represented by a complex landscape of
"hills" and "valleys" located on a square board.  An example of 
such a chaos game representation of a DNA sequence is
presented
in Fig.\ref{fig1} for $6$-tuplets constructed from nucleotides
located in
the first base position in codons, second base position in codons
and third base position in codons in the case of the yeast genes. 
One can observe asymmetric usage of the $6$-words. 
A similar result can be obtained for other values $n$ of length of
 words under consideration.

In general, one can observe many oligomer repeats in the "hills" of the
landscape, especially if one includes noncoding DNA regions. 
Their number is closely related to the mutation
pressure and selection. The statistical properties of
short oligonucleotides have been discussed recently  by Buldyrev et al.
\cite{l_buldyrev2}. In particular, 
they showed that the number of dimeric tandem repeats in
coding DNA sequences is exponential, whereas in noncoding
sequences it is more often described by a power law. 

In the following we restrict ourselves to statistical analysis of 
$6$-tuplets only and to this aim we have considered the
relative frequencies of $6$-words by a simple generalization of the
Eq.\ref{r_Karlin}:

\begin{equation}
z=\frac{P_{ijklmn}}{P_i P_j P_k P_l P_m P_n}
\label{rsingle}
\end{equation}

\noindent
where $P_{ijklmn}$ is the frequency of the word $ijklmn$ in
the genome under consideration,
and $P_i$, $P_j$, \ldots, $P_n$ are the respective nucleotide
frequencies ($i,j,k,l,m,n=A,T,G,C$). 
Thus, $z=1$ means that for a chosen $6$-word the
frequency of its usage in the genome is the same as the expected
probability calculated from the nucleotide occurrence. Both
under-representation and over-representation of $6$-words might introduce
the short-range correlation effects. 
If the words have a biological
sense in DNA texts, they will be correlated at least
in the region of a gene.  

\section{DNA words versus mutations and selection}
The choice of variables $z$ in Eq.\ref{rsingle} to represent effective 
frequencies of $6$-words instead of absolute frequencies
$P_{ijklmn}$ guarantees that trivial correlations, the artefacts 
coming from the nucleotide bias, have been removed. Thus, if the numbers $z$
associated with the respective $6$-words
represent biased random values only,  
their Zipf plot should 
be horizontal. We would like to address the paper by Vandewall and Ausloos
\cite{l_ausloos1} who used this argumentation in their analysis of financial data
- daily fluctuations of the Apple stock price. 

\begin{figure}
\includegraphics[height=8cm]{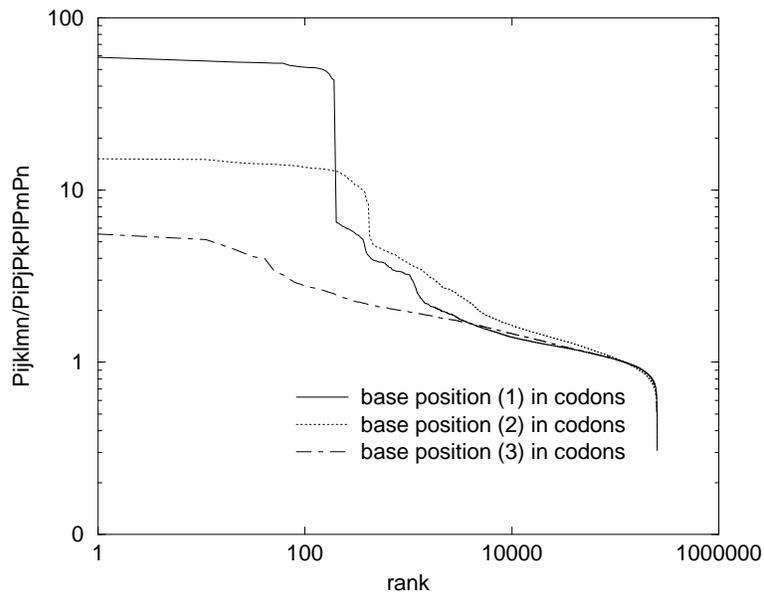}
\caption{Zipf plots for three DNA subsequences originating from
2772 yeast genes, separately for bases in
position (1) in codons, position (2) in codons, and position (3) in codons
in the case when the effective frequencies $z$ of $6$-words were used. 
}
\label{fig2}
\end{figure}

We analyze separately three gene subsequences, 
obtained by splicing nucleotides from position (1) 
in codons, position (2)
in codons and position (3) in codons. Next, the three resulting nucleotide
sequences are partitioned into non-overlapping $6$-tuplets.
Note that some 
$6$-tuplets can be strongly under-represented. The reason is that $6$-tuplets are
already gene-specific and in the extreme case it can happen that a 
$6$-tuplet from a gene under consideration does not appear in any other
genes. This could introduce a strong correlation effect. Therefore, the values
of $z$ in Eq.\ref{rsingle} for a gene under consideration 
are calculated with the help of the frequencies $P_{ijklmn}$, $P_i$, $P_j$, 
\ldots, $P_n$ in a bank of $6$-words representing all genes except the
considered one. 
In Fig.\ref{fig2} we present 
the Zipf plots done for $6$-tuplets 
in the case of 2772 yeast genes taken from
$ftp://genome-ftp.stanford.edu/pub/yeast/genome\_seq/all\_gcg$. 
The results suggest that we can expect non-trivial correlations between
successive $6$-words. The reason for the observed step-like structure 
in Fig.\ref{fig2} is that some deviations of $P_{ijklmn}$ 
from the expected value are more frequent than others and, in general, 
the probability of choosing the next word may depend on several of
the preceding words. 

Note that the representation of genes by the effective frequencies
(Eq.\ref{rsingle}) of their $6$-tuplets loses some 
information concerning base arrangement. 
It is often the case that different $6$-tuplets
have exactly the same  deviation of $P_{ijklmn}$ 
from the expected value in the genome. 
Thus a question could arise: is the $z$ representation
of genes consistent with the L\'{e}vy walk analog of a two-dimensional DNA walk
in space (A-T,G-C), discussed by 
Abramson, Alemany and Cerdeira \cite{l_levy}? 
 In \cite{l_levy} it has been shown that the 
mean square  displacement of the DNA walker follows the power law 
$<r^2(s)> \sim s^{\alpha}$, where $s$ denotes the number of steps and
$\alpha \sim 1.5$ for yeast chromosomes. Once $1 < \alpha <2$
this walk corresponds to the
L\'{e}vy walk.   
We could expect that the distribution of the effective frequencies of $6$-words
should keep the memory of the L\'{e}vy flights performed in space (A-T,G-C).
To show this, it will be convenient for us to introduce a new variable $z'$,
which is the effective frequency $z$ defined in Eq.\ref{rsingle} shifted by 1:

\begin{equation}
z'=z-1 .
\end{equation}

\begin{figure}
\includegraphics[height=8cm]{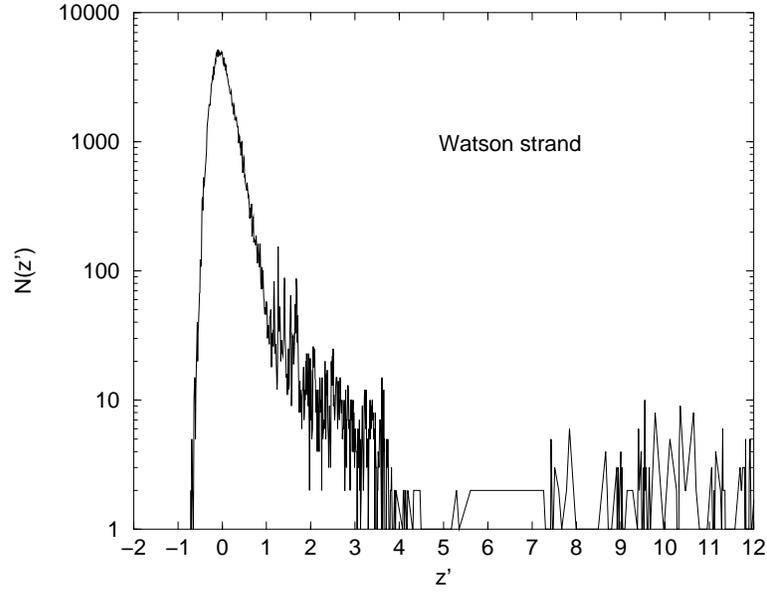}
\caption{Distribution of the numbers 
$z'$ representing   
$6$-words specific for base position (2) in codons of  
 the yeast genes. Here only the genes located at one DNA strand
 have been considered.
}
\label{fig3}
\end{figure}

\begin{figure}
\includegraphics[height=8cm]{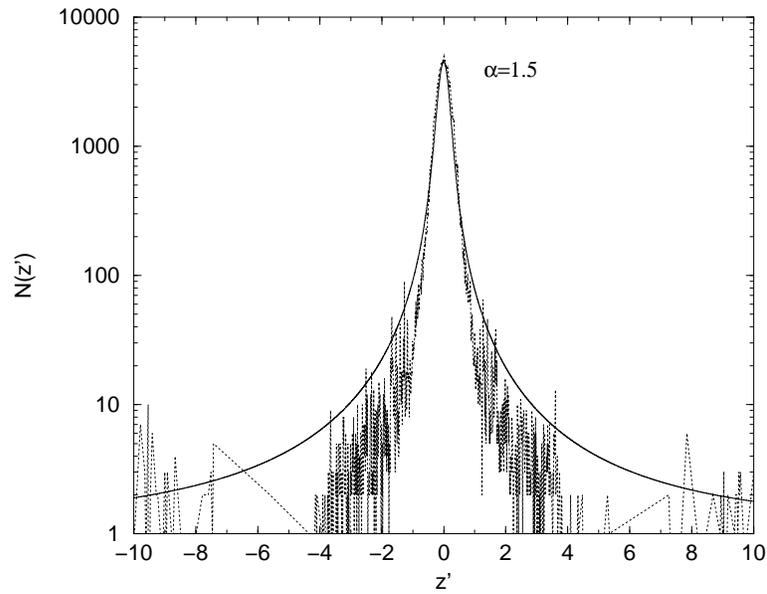}
\caption{Distribution of the numbers $z'$ representing  
$6$-words of yeast genes at base position (2) in codons.
We associated a value $z'$ with genes of one strand, and 
a value $-z'$ with genes of the complementary DNA strand.
The continuous line represents the L\'{e}vy distribution with $\alpha=1.5$.
}
\label{fig4}
\end{figure}

\begin{figure}
\includegraphics[height=8cm]{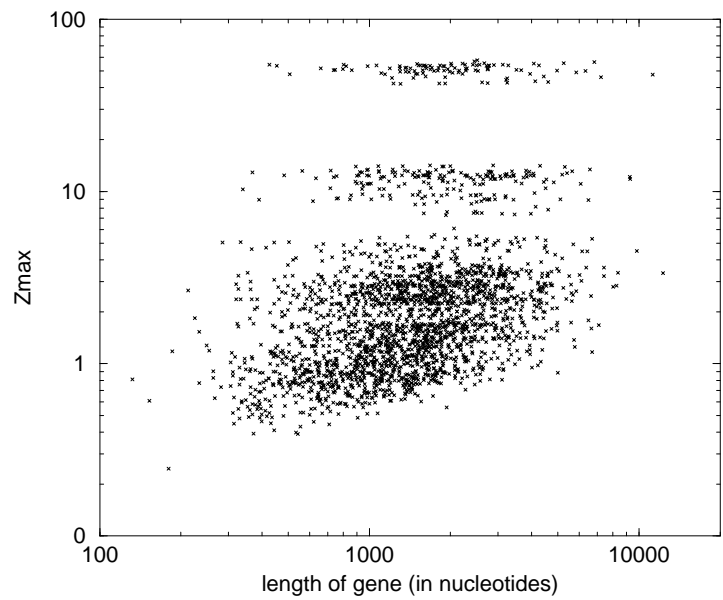}
\caption{For each gene length (in nucleotides), the maximum value of $z$ has been
recorded. 2722 yeast genes have been analyzed.
}
\label{fig5}
\end{figure}

\noindent
In Fig.\ref{fig3} we plotted 
the distribution of numbers $z'$ representing yeast genes solely from one DNA
strand. Almost the same distribution we have got for the numbers $z'$ calculated for genes 
located in the complementary strand. In Fig.\ref{fig4}, we symmetrized the
distribution of the numbers $z'$ by introducing the values $-z'$ in the case of
$6$-tuplets of the complementary DNA strand. The long-tail feature  
of the distribution of the numbers $z'$ is compared in the figure with a 
L\'{e}vy flight
distribution calculated for the exponent $\alpha=1.5$, a value characteristic
for the yeast genome \cite{l_levy}.
The property of the large variance of $z$ is consistent with the suggestion
of non trivial correlation by Fig.\ref{fig2}. 
The results are consistent also
with other data presented in Fig.\ref{fig5}, where for each gene length 
the maximum value of $z'$ has been plotted. In the figure we can observe 
 a trend 
that long genes use more strongly over-represented $6$-words than short genes. 
An analogous situation we have noticed in the case of under-represented $6$-words.
If selection on the extremely over-represented/under-represented $6$-words 
was strong, long genes would be more affected by 
spontaneous mutations 
than short genes. This is suggested also by the results \cite{l_kowal} of  
our Monte Carlo simulations of gene evolution under
constant mutation pressure and selection, where we showed that 
short genes accumulate more
mutations per gene length than the long ones \cite{l_kowal}. 
The fact that spontaneous mutation rates per nucleotide are
inversely correlated with genome size has been first discussed 
by Drake et al. \cite{l_drake} and later by Karlin and Burge
\cite{l_karlin}. Our results might relate this phenomenon 
to the strong over-representation and under-representation of some
$n$-words representing oligonucleotides.

\pagebreak
\begin{flushleft}
{\bf \large  Figure Captions}
\begin{itemize}
\item[Fig.1] The chaos game representation of the  yeast genes 
 in case of
$6$-words constructed from the first base position in codons (left),
second base position in codons (middle), third base position in codons (right). 
\item[Fig.2] Zipf plots for three DNA subsequences originating from
2772 yeast genes, separately for bases in
position (1) in codons, position (2) in codons, and position (3) in codons
in the case when the effective frequencies $z$ of $6$-words were used.
\item[Fig.3] Distribution of the numbers 
$z'$ representing   
$6$-words specific for base position (2) in codons of  
 the yeast genes. Here only the genes located at one DNA strand
 have been considered.
\item[Fig.4] Distribution of the numbers $z'$ representing  
$6$-words of yeast genes at base position (2) in codons.
We associated a value $z'$ with genes of one strand, and 
a value $-z'$ with genes of the complementary DNA strand.
The continuous line represents the L\'{e}vy distribution with $\alpha=1.5$.
\item[Fig.5] For each gene length (in nucleotides), the maximum value of $z$ has been
recorded. 2722 yeast genes have been analyzed.
\end{itemize}
\end{flushleft}

\pagebreak

\end{document}